\documentstyle[aasms4,psfig]{article}

\newcommand{\eq}{\begin{equation}}
\newcommand{\ee}{\end{equation}}

\begin{document}

\title {THERMAL INSTABILITY AND THE FORMATION OF CLUMPY GAS CLOUDS}

\author {A. Burkert\altaffilmark{1} and D. N. C. Lin\altaffilmark{2}}

\altaffiltext
{1}{Max-Planck-Institut f\"ur Astronomie, K\"onigstuhl 17,
D-69117 Heidelberg, Germany.
E--mail: burkert@mpia-hd.mpg.de}

\altaffiltext
{2} {UCO/Lick Observatory, University of California, Santa Cruz, California, 
95064.
E--mail: lin@lick.ucolick.org}
 
\vspace{3cm}
\slugcomment{{\em Astrophysical Journal, in press}}

\begin{abstract}
The radiative cooling of optically thin gaseous regions and the
formation of a two-phase medium and of cold gas clouds with a clumpy
substructure is investigated.  We demonstrate how clumpiness can
emerge as a result of thermal instability. In optically thin clouds,
the growth rate of small density perturbations is independent of their
length scale as long as the perturbations can adjust to an isobaric
state. However, the growth of a perturbation is limited by its
transition from isobaric to isochoric cooling when the cooling time
scale is reduced below the sound crossing time scale across its length
scale. The temperature at which this transition
occurs decreases with the length scale of the perturbation.
Consequently small scale perturbations have the potential
to reach higher amplitudes than large scale perturbations.  When
the amplitude becomes nonlinear, advection overtakes the pressure gradient
in promoting the compression  resulting in an accelerated growth of the
disturbance.  The critical temperature for transition depends on
the initial amplitude.  The fluctuations which can first reach
nonlinearity before their isobaric to isochoric transition will
determine the characteristic size and mass of the cold dense clumps
which would emerge from the cooling of an initially nearly homogeneous
region of gas. Thermal conduction is in general very efficient in erasing
isobaric, small-scale fluctuations, suppressing a cooling instability.
A weak, tangled magnetic field can however reduce
the conductive heat flux enough for low-amplitude fluctuations to grow
isobarically and become non-linear if their length scales are of order $10^{-2}$ pc. 
If the amplitude of the initial perturbations is a decreasing function 
of the wavelength, the size of the emerging clumps will
decrease with increasing magnetic field strength.
Finally, we demonstrate how a 2-phase medium, with cold clumps being pressure confined
in a diffuse hot residual background component, would be sustained if
there is adequate heating to compensate the energy loss.
\end{abstract}


\section{Introduction}

The interstellar medium and cold gas clouds are characterized by a
clumpy substructure and a turbulent velocity field (Larson 1981, 
Blitz 1993).  As
molecular clouds are the sites of star formation, their formation,
internal structure and dynamics determines the rate of star formation
and the properties of young stars, such as their mass function or
binarity.  The understanding of the origin of cold clouds and
their internal substructure is therefore of fundamental importance for
a consistent theory of star formation and galactic evolution.

In the nearby clouds, the dispersion velocity inferred from molecular
line width is often larger than the gas sound speed inferred from
the line transition temperatures (Solomon et al 1987). 
MHD turbulence may be responsible for the stirring of these clouds
(Arons \& Max 1975).  This conjecture is supported by the polarization maps
and direct measurements of field strength in some star forming regions
(Myers \& Goodman 1988, Crutcher et al. 1993). 
Recent simulations of MHD turbulence, however, suggest
that it dissipates rapidly (Gammie \& Ostriker 1996, MacLow et al 1998, MacLow 1999,
Ostriker et al. 1999).  One possible
source of energy supply is winds and outflows from young stellar objects 
(Franco \& Cox 1983, McKee 1989).  But in regions where star formation is inactive, clumpy
structure with velocity dispersion is also observed.  Thus,
the origin and energy supply of clumpy cloud structure remains an
outstanding issue.

On small scales, magnetic field pressure is important in regulating
infall and collapse of protostellar clouds and the formation of low-mass
stars (Mouschovias \& Spitzer 1976, Nakano 1979, Shu 1993). 
For clouds with sub-critical masses, 
gravitational contraction is proceeded by ambipolar diffusion which 
for typical cloud densities operates on a timescale $\tau_B \sim 10^{7-8}$ yr
(Lizano \& Shu 1989, Mouschovias 1991). 
In regions with intense star formation activities such as
the central region of Orion, $\tau_B$ for individual dense clumps
is comparable to the typical age of the young stellar objects.  But,
the spread in stellar ages ($\Delta \tau_\ast \sim 10^6$ yrs)
appears to be considerably shorter than $\tau_B$ (Carpenter et al. 1997, Hillenbrand, 1997).
This coeval star formation history requires either
a coordinated trigger mechanism for star formation
within initially magnetically supported clumps or subcritical 
collapse, fragmentation and star formation of a larger molecular cloud region
in which the magnetic field plays a weak role.

A rapid and coordinated episode of star formation can also be inferred
in globular clusters (Brown et al. 1991, 1995, 
Murray \& Lin 1992, Lin \& Murray 1992).  In some metal deficient 
clusters such as M92, the total amount of heavy elements corresponds
to the yield of a few supernovae.  If star formation has proceeded
over a duration $\Delta \tau_\ast$ comparable to the expected life span ($\sim$ a few 
$10^6$ yrs) of massive stars, a significant metallicity spread 
would be expected, in contrast to the observations (e.g. Kraft 1979).
At least in these systems, $\Delta \tau_\ast <\tau_B$ and star 
formation may have proceeded through supercritical collapse.  The
dynamical timescale of most clusters at their half mass radius is 
$\tau_d \approx 10^6$ yr.  Any energy dissipation associated with
the episode of star formation would imply an even longer dynamical timescale 
in the proto cluster cloud prior to that event.  We infer that 
$\Delta \tau_{\ast}$ was comparable to or shorter than the dynamical 
timescale of the proto cluster clouds indicating a rapid fragmentation
and star formation episode.

In this paper, we focus on the rapid emergence of clumpy structure during
the formation and collapse of a thermally unstable supercritical cloud. This process
is relevant to the formation of stellar clusters as well as galaxies.
We assume that the clouds condense out of a diffuse hot medium
as a result of thermal instability.
Large condensations with cooling timescales $\tau_c > \tau_d$ are thermally stable
because they can adjust through contraction such that their radiative
losses may be compensated by the release of their gravitational energy.
Runaway cooling of the gas through thermal instability however
occurs in clouds with $\tau_c < \tau_d$.  
In order to form clumps within an initially almost homogeneous
cloud, internal density fluctuations must grow rapidly on a timescale
short compared to the mean dynamical timescale of the entire cloud.
Small scale density fluctuations would begin to dominate if either the
growth timescale or the limiting amplitude is a decreasing function of
the perturbations' length scale.  One possible fragmentation
mechanism is gravitational collapse. The reduction in the cloud's
temperature reduces its Jeans' mass, leading to the onset of
gravitational instability and collapse.  However, for a non rotating,
cold, homogeneous gaseous region, gravitational instability alone
cannot induce fragmentation
because the growth rate is essentially independent of length scale
such that the growth timescale for the density contrast is comparable
to the dynamical timescale of the whole cloud (Hunter 1962).  This has also been
shown by numerical collapse simulations of initially gravitationally
unstable perturbed gas clouds (e.g. Burkert \& Bodenheimer 1993, 1996,
Burkert, Bate \& Bodenheimer 1997).  If the initial density
perturbations $\delta_0$ are linear ($\delta_0 < 1$), fragmentation is
suppressed until the gas cloud has collapsed into either a disk or a
dense filamentary substructure.

We propose that clumpyness in clouds arises naturally from their
formation through a cooling instability which acts on timescales that can
be much shorter than the dynamical timescale of the cloud. 
In a pioneering paper, Field
(1965) derived a criterion for a cooling gas to be unstable to the
growth of thermal condensations. He showed that thermal instability
can lead to the rapid growth of density perturbations from
infinitesimal $\delta_0$ to nonlinear amplitudes on a cooling
timescale $\tau_c$ which for typical conditions in the interstellar
medium is short compared to the dynamical timescale.  If $\tau_c$
increases with decreasing density any small density difference would
induce a temperature difference between the cooler perturbed region
and the warmer background. Across the interface between the two-phase
medium, differential cooling leads to a pressure gradient which
induces a gas flow from the lower-density background towards the
higher-density perturbed region.  The density enhancement in the
cooler region further reduces its cooling timescale compared to that
of the background where $\tau_c$ increases. A more detailed
investigation of the growth of condensations in cooling regions has
been presented by Schwarz et al. (1972) who included also the effects
of ionization and recombination and by Balbus (1986) who examined the
effect of magnetic fields. The classical model of the interstellar
medium where heating balances cooling was presented by Field et
al. (1969).  A recent progress report on the theory of thermal
instability is given by Balbus (1995).

Although thermal instability proceeds faster than the collapse of the
cloud, its growth rate is determined by the local cooling rate.
During the initial linear evolution, variations in the initial over density
(or under temperature) might lead only to a weak dependence of the growth
timescale on the wavelength.
In this paper we show however that there 
exist two important transitions which are
very sensitively determined by the wavelengths of perturbations. 1) The growth
of a perturbation is limited by its transition from isobaric to
isochoric cooling, when the cooling time scale is reduced below the
sound crossing time scale across the wavelength of the perturbation.
This transition occurs at a lower temperature, with correspondingly
larger over density, for perturbations with smaller wavelengths.
2) For those perturbation which can become nonlinear before the isobaric to
isochoric transition, advection overtakes the pressure gradient in
promoting the compression and growth of the perturbed region at an
accelerated rate.  The fluctuations which can first reach nonlinearity
would dominate the growth of all perturbations with
longer wavelengths and homogenize disturbances with smaller
wavelengths.  Thus, they determine the characteristic size and mass of
the cold dense clumps which would emerge from the cooling of an
initially nearly homogeneous cloud. Thermal conduction could in general
erase these fluctuations, suppressing the instability. Weak, tangled
magnetic fields would however be efficient enough in reducing the
conductive flux, allowing the medium to break up into cold clumps
on the characteristic length scale.

We study the cooling
and fragmentation of gas using simplified power-law cooling functions.
Since we are primarily interested in supercritical clouds, we neglect
the effect of magnetic fields.  Note that even a weak magnetic field
could have an important destabilizing influence in thermal instability
(Loewenstein 1990, Balbus 1995). In \S2, we obtain approximate analytic
solutions which describe the evolution of a linear density
perturbation in the isobaric and nearly isochoric regime.  We show
that the growth of over density in a thermally unstable fluctuation is
limited by a transition from isobaric to isochoric evolution and that
the limiting amplitude is a decreasing function of the length scale.
We verify our analytic approximations with numerical, hydrodynamical
calculations which are also used in \S3 to study the transition into
the non-linear regime.  In \S4 we investigate the cooling of
interacting perturbations and determine the critical length scale of
clumps that emerge through thermal instability.  The importance of
thermal conduction is investigated in \S5. In \S6 we discuss the affect of
heating processes and the formation of a stable 2-phase medium.
Finally, we summarize our results and discuss
their implications in \S7.

\section{The Initial Evolution of Thermal Instability}

The dynamical evolution of the gas is described by the hydrodynamical
equations

\begin{equation}
\frac{\partial \rho}{\partial t} + \sum_{k=1}^3 
\frac{\partial \rho  U_k}{\partial x_k} = 0
\end{equation}
\begin{equation}
\frac{\partial U_j}{\partial t} + \sum_{k=1}^3U_k 
\frac{\partial U_j}{\partial x_k} + \frac{R_g}{\mu \rho}
\frac{\partial}{\partial x_j} \left(\rho T \right) = 0
\end{equation}
\begin{equation}
\frac{\partial T}{\partial t} + \sum_{k=1}^3 U_k 
\frac{\partial T}{\partial x_k} + (\Gamma - 1)T \sum_{k=1}^3 
\frac{\partial U_k}{\partial x_k} = -\frac{\rho \Lambda}{C_v}
\end{equation}
where j=1,2,3 is the coordinate index, $C_v = R_g/\mu (\Gamma -1)$ is
the heat capacity, $R_g, \mu$ and $\Gamma$ are the gas constant, mean
molecular weight and adiabatic index, respectively.

In the unperturbed state, the gas remains at rest ($U_j=0$) and its
density attains a constant value, $\rho_0$.  The time dependent energy
equation (3) gives
\begin{equation}
C_v \frac{\partial T_0}{\partial t} = - \rho_0 \Lambda
\end{equation}
where the cooling rate $\Lambda= \Lambda_0 T_0 ^\beta$. The power
index is determined by the detailed atomic processes.  Since we are
primarily interested in the physical evolution of thermal instability,
we adopt a simple constant $\beta$ prescription.  The cooling would be
thermally unstable (with $\tau_c = T_0/\partial T_0/\partial t$ as an
increasing function of $T_0$) in the isochoric region if $\beta < 1$
and in the isobaric region if $\beta < 2$.  In the absence of external
heating, the unperturbed gas temperature $T_0$ can be expressed as a
function of the dimensionless time variable $\tau \equiv t /
\tau_c(0)$ such that
\begin{equation}
T_0(t)=T_0(0) \left(1 - (1-\beta) \tau \right)^{\frac{1}{1-\beta}}
\end{equation}
where $T_0(0)$ and $\tau_c (0)\equiv C_v / \rho_0 \Lambda_0 T_0(0)
^{\beta-1}$ are the initial (at $t=0$) temperature and cooling
timescale, respectively.

\subsection{The Perturbed Quantities}

The evolution of the perturbed density ($\rho_1=\rho-\rho_0$),
temperature ($T_1=T-T_0$), and velocities ($U_j$) are derived from the
linearization of the equations (1) to (3):
 \begin{equation}
\frac{\partial}{\partial t} \frac{ \rho_1}{\rho_0} 
= -\sum_{j=1} ^3 \frac{\partial U_j}{\partial x_j},
\label{a6}
\end{equation}
\begin{equation}
\frac{\partial U_j}{\partial t} = - \frac{R_g T_0}{\mu} \frac{\partial 
}{\partial x_j} \left( \frac{T_1}{T_0} + \frac{\rho_1}{\rho_0} \right),
\label{a7}
\end{equation}
and
\begin{equation}
\frac{\partial}{\partial t} \frac{T_1}{T_0} = -(\Gamma -1) 
\sum_{j=1} ^3 \frac{\partial U_j}{\partial x_j} -\frac{1}{\tau_c}
\left( \frac{\rho_1}{\rho_0} + (\beta-1) \frac{T_1}{T_0} \right)
\label{a8}
\end{equation}
where
$$\tau_c (t) \equiv - \frac{T_0}{d T_0 / dt} = \tau_c(0) - 
(1-\beta)t$$ is the 
characteristic cooling timescale at the instant of time t.

Since the perturbation equations are linear in $x_j$, we adopt a local
approximation in which the positional dependence of all the perturbed 
quantities is proportional to exp(i$k_j x_j$) where $k_j$ is the wave
number in the $j$th direction.  Substituting a dimensionless velocity 
variable $V_j = i k_j \tau_c(0) U_j$, the perturbed equations reduce to
\begin{equation}
\frac{\partial}{\partial \tau} \frac{ \rho_1}{\rho_0} = -\sum_{j=1} ^3 V_j,
\label{a10}
\end{equation}
\begin{equation}
\frac{\partial V_j}{\partial \tau} =  K_j ^2 \left( 1 - (1 - \beta) \tau
\right) ^{1 \over 1 - \beta}  \left( \frac{P_1}{ P_0} \right),
\label{a11}
\end{equation}
where $\frac{P_1}{ P_0} = \frac{T_1}{T_0} + \frac{\rho_1}{\rho_0}$ is
the perturbed pressure, $K_j \equiv \tau_c(0) k_j \sqrt {R_g T_0
(0)/\mu}$ is the ratio of the initial cooling to sound crossing
timescale over a characteristic wavelength $2 \pi/ k_j$, and
\begin{equation}
\frac{\partial}{\partial \tau} \frac{T_1 }{T_0} = - (\Gamma -1) 
\sum_{j=1} ^3 V_j -\frac{\tau_c (0)}{ \tau_c}
\left( \frac{\rho_1}{\rho_0} + (\beta-1) \frac{T_1}{T_0} \right).
\label{a12}
\end{equation}
For a perfect gas, the unperturbed pressure $P_0=R_g \rho_0 T_0/\mu$
decreases at the same rate everywhere.  We find from Eqs (\ref{a10})
and (\ref{a12}) that the amplitude of the perturbed pressure is
\begin{equation}
\frac{\partial}{\partial \tau} \frac{ P_1}{P_0}= 
\frac{\partial}{\partial \tau} \left( \frac{ \rho_1}{\rho_0} 
+ \frac{T_1}{T_0} \right) = 
- \Gamma \sum_{j=1} ^3 V_j - \frac{1}{1-(1-\beta)\tau} 
\left((2-\beta)\frac{\rho_1}{\rho_0} - (1-\beta)\frac{P_1}{P_0} \right).
\label{a13}
\end{equation}

\subsection{The initially isochoric regime with K $\leq$ 1}

For computational simplicity, we now consider a 1-D limit treatment in
which the initial (at $\tau=0$) amplitude of $\rho_1$ equals to a
finite value $\rho_a$ with that of $V_1$ and $P_1$ equal to zero.
These conditions correspond to an initially almost homogeneous, hot
region of gas in pressure equilibrium.  To third order in $\tau$ the
eqs (\ref{a10}), (\ref{a11}), and (\ref{a13}) give the following
approximate solution

\begin{equation}
\frac{\rho_1}{\rho_0} \simeq \frac{\rho_a}{\rho_0}\left(1 
+ \frac{K^2}{6}\left(2-\beta\right) \tau^3\right)
\end{equation}
\begin{equation}
V \simeq \frac{(\beta-2)K^2}{6}\frac{\rho_a}{\rho_0}
\left(3\tau^2 + 2\beta\tau^3\right)
\end{equation}
\begin{equation}
P \simeq (\beta-2)\frac{\rho_a}{\rho_0}\left(\tau^2 
+ \left(1-\beta\right)\tau^3 + \left(\frac{4}{3} - 
\frac{7}{3}\beta + \beta^2-\frac{\Gamma K^2}{6}\right)\tau^3\right)
\end{equation}

Figure 1 compares this solution with a numerical integration of the
complete non-linear hydrodynamical equations (1) to (3) for K=1 and
K=0.5.  We use a 1-dimensional version of the second-order Eulerian
hydro code which is described in Burkert and Bodenheimer (1993).  The
agreement between the numerical results (solid lines) and analytical
solution (dots) is excellent, even for large values of $\tau \approx 
1$ where the basis of the analytic approximation is no longer valid.

Due to slightly more efficient cooling within the density perturbation
a small pressure gradient builds up. In an attempt to maintain
pressure balance, the slightly warmer gas in the low-density regions
continually compresses the more dense and cooler parts. Consequently
the over-density in the perturbed region increases as the gas
cools. Figure 1 and the equations (13) to (15) show that for small
values of $K \leq 1$ the growth rate of the density enhancement
depends on the size of the perturbation and increases with increasing
values of K or decreasing wavelength. However, due to its long sound
crossing timescale, the perturbation cannot be compressed
significantly while cooling; it cools almost isochorically (Parker
1953). After one cooling timescale the gas temperature has reached its
minimum value with the density enhancement still in the linear regime.
Now, the pressure gradient reverses, erasing the fluctuation.

\subsection{The initially isobaric regime: K $\geq$ 1}

The solid lines in figure 2 show a numerical calculation of the
evolution of a density perturbation with K = 200, $\beta = 0$ and
$\Gamma = 5/3$.  For $K >> 1$, perturbations can react quickly on any
pressure gradients due to the short sound crossing timescale, relative
to the cooling timescale.  The simulations indicate a solution which
consists of a fast oscillatory part and a slowly growing part.
Linearizing the slowly growing part, we find from the equations (9) to
(12) the following approximate solution:

\begin{equation}
\frac{\rho_1}{\rho_0} \simeq \frac{\rho_a}{\rho_0}
\left(\frac{i \omega \Gamma}{i \omega \Gamma -2 + \beta}\right)
\left( 1 + \frac{2 - \beta}{\Gamma} \left( \tau 
+ \frac{i}{\omega} e^{i \omega \tau} \right) \right),
\label{a20}
\end{equation} 
\begin{equation}
V\simeq \frac{\beta-2}{\Gamma} \frac{\rho_a}{\rho_0} 
\left(\frac{i \omega \Gamma}{i \omega \Gamma -2 + \beta}\right)
\left( 1 + i \omega \tau - e^{i \omega \tau} \right), 
\label{a21}
\end{equation} 
and
\begin{equation}
\frac{P_1}{P_0} \simeq \frac{(2- \beta) i \omega }{\Gamma K^2}
\frac{\rho_a}{\rho_0} 
\left(\frac{i \omega \Gamma}{i \omega \Gamma -2 + \beta}\right)
\left( 1 + (1 - \beta) \tau - e^{i \omega \tau} \right).
\label{a22}
\end{equation} 

The characteristic frequency is determined by a cubic dispersion relation
\begin{equation}
\omega^3 + i (1-\beta) \omega^2 - K^2 \Gamma \omega -i(2-\beta) K^2 =0.
\label{a23}
\end{equation}

\noindent A similar relation was discussed by Balbus (1995).  Note
that the linearized solution is also valid for $\tau > 1/\omega$ as
long as $\tau << 1$.  As $K >> 1$ the two dominant real roots ($\omega
\approx \pm \sqrt{\Gamma} K$) of the dispersion eq(\ref{a23}) yield
oscillatory parts in $\rho_1/\rho_0$, $V$, and $P_1/P_0$.  The upper
panels of figure 2 show that for $\tau < 0.1$ the analytical
approximation is in good agreement with the numerical integration of
the nonlinear hydrodynamical equations.

For all length scales, the ratio of sound propagation to cooling timescale 
\begin{equation}
Q \equiv \tau_c (t) k \sqrt{R_g T_0 (t) /\mu} 
= K (1-(1 - \beta)\tau)^{3-2 \beta \over 2- 2 \beta} 
\label{a24}
\end{equation}

\noindent decreases during the subsequent evolution.  Provided $Q > >
1$, eq(\ref{a22}) implies that the magnitude of $P_1/P_0$ is much
smaller than both $V$ and $\rho_1/\rho_0$. That is, the fluctuation
reacts isobaric.  Adopting $P_1/P_0 \approx 0$ and neglecting the
oscillatory term, the equations (9) and (12) can be combined to

\begin{equation}
-V = \frac{\partial}{\partial \tau}\left(\frac{\rho_1}{\rho_0}\right)
= \frac{(2-\beta)}{\Gamma(1-(1-\beta)\tau)}\frac{\rho_1}{\rho_0}
\end{equation}
\noindent with the solution 
\begin{equation}
\frac{\rho_1}{\rho_0} = \frac{\rho_a}{\rho_0} \left( 1 - (1 -\beta)
\tau \right) ^{- \frac{ 2 - \beta} {(1 - \beta) \Gamma}} 
\label{a25}
\end{equation}
and 
\begin{equation}
V =- \frac{(2 - \beta)}{\Gamma} \frac{\rho_a}{\rho_0} 
(1 - (1-\beta)\tau) ^{- \frac{ 2 - \beta} {(1 - \beta) \Gamma} -1}.
\label{a26}
\end{equation} 

In contrast to fluctuations with $K < 1$ the evolution of isobaric
fluctuations is independent of K. For $(1-\beta)^{-1} > > \tau >
\omega^{-1}$, solutions for $\rho_1 /\rho_0$ in Eqs (\ref{a20}) and
(\ref{a25}) are in agreement to first order in $\tau$. The lower
panels of figure 2 show that within a cooling time both
$\rho_1/\rho_0$ and $V$ are amplified to very large values.  The
agreement between the numerical calculation and the analytical
prediction (equation 22 and 23) is excellent.  The opposite signs of
$\rho_1/\rho_0$ and V confirm that mass is being pushed into the cool
dense regions.  For $\tau \approx 1$, the numerically derived density
enhancement falls below the predicted values as the fluctuation
becomes isochoric and contributions from the perturbed pressure 
cannot be neglected anymore.

\subsection{Transition to Isochoric Evolution and the Emergence 
of Small Scale Perturbations}

Figure 3 shows the density evolution of initially isobaric
fluctuations with different ratios of cooling to sound crossing times
K as determined from the numerical calculations.  The initially
isobaric growth of the density fluctuations is independent of
wavelength and K and in excellent agreement with equation (22) (dashed
curve).  The perturbations transform  however to the isochoric
solution for the epoch after $Q$ has declined below unity.
Thereafter, gas in the perturbed region cools off faster than it
can adjust to a pressure equilibrium with the surrounding region.
Subsequently, the over density of the perturbed region is slowly
modified by the inertial motion $V_{trans}$ of the gas at the time of
the transition and its growth stalls.  In Figure 3 the transition into
the isochoric regime is indicated by the overdensity falling below the
expected value shown by the dashed thick line.

Although the growth of the perturbed quantities does not explicitly
depend on the wavelength $k$, the critical transition time when $Q
\approx 1$
\begin{equation}
\tau_{trans} = \frac{1-K^{\frac{2 \beta-2}{3 -2 \beta}}}{1-\beta }
\label{a32}
\end{equation}
is a function of $K$ (and $k$).  At this transition point, the over 
density in the perturbed region is

\begin{equation}
\frac{\rho_{trans}}{\rho_0} = \frac{\rho_a}{\rho_0} K^{\frac{(4 - 2
\beta)}{(3 - 2 \beta) \Gamma}}.
\label{a33}
\end{equation}

\noindent and the velocity is

\begin{equation}
V_{trans} = - \frac{2-\beta}{\Gamma}\frac{\rho_a}{\rho_0} 
K^{(\frac{2-2\beta}{3-2\beta})(1+\frac{2-\beta}{(1 - \beta) \Gamma})} .
\end{equation}

\noindent In the isochoric regime the amplitude of the perturbed
density increases as

\begin{equation}
\frac{\rho_1}{\rho_0} = \frac{\rho_{trans}}{\rho_0} 
-V_{trans}*(\tau -\tau_{trans})
\end{equation}

\noindent which increases much less steeply than the isobaric 
fluctuations (equation 22).

For thermally unstable clouds, $\beta < 1$ such that $\rho_{trans} /
\rho_0$ is an increasing function of K or a decreasing function of the
wavelength ($\lambda$) of the perturbations.  Despite the independence
of the rate of change of $\rho_1/\rho_0$ on $\lambda$, equation 24
shows that for $\beta < 1$ the short length scale disturbances undergo
isobaric to isochoric transition at a later time and therefore acquire
a greater limiting amplitude than the long length scale disturbances.
Thus, the short length scale disturbances would emerge to dominate the
structure of the cloud unless the initial perturbation amplitude
$\rho_a/\rho_0$ increases with $K$ more rapidly than $K ^{(2 \beta
-4)/(3 - 2 \beta) \Gamma}$.  This evolution is physically equivalent
to the fragmentation process in which the contrast between the
enhanced density in a disturbance and the average cloud density
becomes most pronounced on the smallest scales.

\section{The Transition into the Nonlinear Regime}
       
In Fig. 3 fluctuations with very large values of K show yet another
evolution: for later times the overdensity rises faster than predicted
by equation (22). These fluctuations become nonlinear with
$\rho_1/\rho_0 > 1$ before the transition into the isochoric
regime. The critical value of K for this evolution can be estimated
from equation (25) assuming $\rho_{trans}/\rho_0 = 1$:

\begin{equation}
K_{crit} = \left( \frac{\rho_a}{\rho_0} \right)
^{\frac{(2\beta - 3)\Gamma }{4-2\beta}}
\label{a28}
\end{equation}

For $K > K_{crit}$ the analytical approximations discussed previously
are not valid anymore and we have to investigate the evolution
numerically, solving the complete non-linear hydrodynamical
equations. The simulations shown in figure 3 assumed $\beta = 0$,
$\Gamma = 5/3$ and $\rho_a/\rho_0=10^{-3}$. For these values the
simple approximation (28) predicts $K_{crit}=5600$ which is roughly in
agreement with the numerical results where the transition into the
nonlinear regime occurs more smoothly between K=1000 and
K=5000. Equation (28) somewhat overestimates $K_{crit}$ because
nonlinear effects actually become important earlier, when the
overdensity is in the range $0.1 < \rho_1/\rho_0 < 1$.

Figure 4 shows the structure and evolution of a non-linear
fluctuation. During the early isobaric evolution the pressure gradient
(lower right panel) is negligible. A small pressure gradient builds up
in the nonlinear regime, where the profiles cannot be approximated
anymore by sinusoidal functions but instead become strongly peaked
towards the center. Non-linear fluctuations grow fast with the density
and temperature reaching their maximum and minimum values,
respectively, at a time $\tau_{crit} < 1 $ which is shorter than a
cooling time.  Due to the fast growth in the nonlinear regime,
$\tau_{crit}$ is roughly given by the time when $\rho_1/\rho_0 = 1$.
From Eq (\ref{a25}), we find 
\begin{equation}
\tau_{crit} = \frac{1}{1-\beta} \left( 1 - \left( \frac{\rho_a}{\rho_0} 
\right)^{\frac{(1-\beta )\Gamma}{2-\beta}} \right).
\end{equation}
At $t= \tau_{crit}$, the dimensionless velocity  (see Eq. 
\ref{a26})
\begin{equation}
V_{crit} = V(\tau_{crit}) 
= \frac{\beta-2}{\Gamma} \left( \frac{\rho_a}{\rho_0} \right)^{
-\Gamma \frac{1-\beta}{2-\beta}}
\end{equation}
\noindent is much larger than unity
for perturbations with small initial amplitudes such that
contributions due to nonlinear advection (such as $U_j \partial \rho /
\partial x_j$, $U_j \partial U_j / \partial x_j$, and $U_j \partial T
/ \partial x_j$) would exceed the linear contributions contained in
the perturbed equations (\ref{a6}-\ref{a8}) before the over density
$\rho_1$ has become comparable to $\rho_0$ (see above).  Advection
generally enhances the effect of compression and promotes the growth
of density contrast at an accelerated rate.

Although the time of maximum compression for a fluctuation with $K >
K_{crit}$ does not depend explicitly on the length scale, it is
determined by the initial amplitude $\rho_a/\rho_0$ of the
perturbation which may be a function of the wavelength. For an
initial power-law perturbation in which $\rho_a/\rho_0 = A_0 (k/k_0)
^\eta$, 
\begin{equation}
\tau_{crit} = \frac{1}{1-\beta} \left( 1 - \left( A_0^{\frac{1}{\eta}} \frac{k}{k_0} 
\right)^{\frac{(1-\beta )\Gamma \eta}{2-\beta}} \right).
\end{equation}

If the amplitude of the initial perturbation is an increasing function of
the wavelength (which corresponds to a negative $\eta$),
$\tau_{crit}$ would be an increasing function of $k$ in the thermally
unstable region with $\beta <1$.  In this case, nonlinearity would be
first reached on the largest length scale with $K > K_{crit}$.  If,
however, the amplitude of the initial perturbation is a decreasing
function of the wavelength ({\it i.e.} $\eta >0$), 
$\tau_{crit}$ would be a decreasing
function of $k$ and nonlinearity would be reached on the smallest
scale first.  Note that for $1 < \beta < 2$, the dependence of 
$\tau_{crit}$ on $\eta$ and $k$ is reversed.

In Figure 3, a temperature independent cooling function has been
used. In order to determine the dependence on the specific
form of the cooling function, additional simulations have been performed,
adopting a more realistic cooling function (Dalgarno \& McCray 1972) which assumes
solar element abundance and collisional equilibrium ionization.
Note that for temperatures T $> 10^4$ K the cooling rate is several
orders of magnitudes larger than for T $< 10^4$ K, defining two
different temperature regimes with very different cooling timescales.
The simulations show that the previous results remain
valid for each of these temperature regimes. Starting in the low-temperature
regime, a fluctuation will become non-linear for K $>$ K$_{crit}$ and cool
down to the lowest allowed temperature. The same is true for fluctuations
that start in the high-temperature regime. Non-linear fluctuations in
this regime do however stop cooling efficiently at T $\approx 10^4$ K,
leading to high-density clumps with such a temperature.

\section{Interacting Fluctuations and the Emergence of 
Substructure with a Critical Wave Length}

Up to now we have investigated the evolution of isolated fluctuations.
In reality however a cooling gaseous region consists of a
superposition of fluctuations with different wavelengths and
amplitudes.  As we indicated in the previous section, the outcome of
the thermal instability may be determined by the wavelength dependence in the
initial amplitude of the perturbations.

In order to illustrate various competing effects such as isobaric to
isochoric transition and the onset of nonlinear growth, we present in
Figure 5 a series of models with $\beta=0$ and $\Gamma = 5/3$, where
the initial density distribution consists of the superposition
of two fluctuations with ratios of wavelengths $\lambda_1/ \lambda_2
= 20$ and amplitude ratios $\rho_{a,1}/ \rho_{a,2}= 2$ which
corresponds to $\eta=-0.23$. Four values of $\lambda_1$ were
chosen and they correspond to $K_1$=1,10,100 and 1000, respectively.
The $K_2$ values for the smaller perturbation are always a factor 20
larger.  Since the initial overdensity $\rho_{a,1}/\rho_0 =0.01$,
the critical value of K for nonlinear evolution is $K_{crit}=316$, according to
equation (\ref{a28}).  We show the density distribution after 1 $\tau_c(0)$.  In
the upper left panels of Fig. 5, the fluctuations have values of $K_1=1$ and
$K_2=20$ which are small compared to $K_{crit}$. Their growth
therefore stalls due to transition into the isochoric regime and the
overdensity after a cooling time is still linear. The smaller
fluctuation dominates at the end because its isochoric transition
occurs later and at a higher overdensity than for the larger
perturbation. In the upper right panels with $K_1=10$ and $K_2=200$
the smaller perturbation is again dominating after $\tau=1$ although,
now, the density distribution is also affected by the underlying
larger perturbation.  In both cases the density within the density
peaks does not decrease much with respect to its initial value. The
situation is different in the lower left panel with $K_1=100$ and
$K_2=2000$.  Here the smaller perturbation has become nonlinear,
generating small dense clumps which stand out against the larger
perturbation.  Up to now, the smaller perturbation was always
dominating the density distribution after a cooling time. The
situation is however different in the lower right panel where 
$K_1 > K_{crit}$. Now, the larger perturbation becomes nonlinear and
advection drives all the gas and its small scale fluctuations into one
very dense, cold clump that is embedded in a hot diffuse environment,
erasing smaller scale fluctuations.

The dependence of structure formation on the initial power-law perturbation index
$\eta$ is illustrated in figure 6 which shows the initial  and final density distribution
of two interacting perturbations with $\lambda_1/\lambda_2 = 10$, $K_1 = 2 \times 10^4$,
$\rho_{a,1}/\rho_{a,2}=10$ ($\eta=-1$) in the left panels and
$\rho_{a,1}/\rho_{a,2}=0.1$ ($\eta=1$) in the right panels.
As expected, in the case of $\eta=-1$ the larger perturbation becomes nonlinear first,
leading to one massive density peak  after a cooling time. For $\eta=1$, the small
length scale perturbations begin to dominate after a cooling time, breaking the region up into
dense clumps on the smallest scale. 
More specifically, if the amplitude
of the initial perturbation increases with increasing wavelength ($\eta < 0$),
clumps will form with length scales $\lambda \approx \lambda_{crit}$.
Otherwise, the sizes of the fastest growing perturbations will be
$\lambda \approx \lambda_{\kappa}$. In this case, the clump sizes should
decrease with increasing magnetic field strength.

\section{The importance of thermal conduction}
During the growth of linear density perturbations in the isobaric
regime the resulting temperature gradient will induce conductive
heating of the fluctuations.

\subsection{Thermal conduction in the absence of magnetic fields}

Several studies (e.g. 
McKee \& Begelman 1990, Ferrara \& Shchekinov 1993) have demonstrated
that thermal conduction could stabilize and even erase a density 
perturbation if its scale is smaller than the Field length (Field 1965)

\begin{equation}
\lambda_F = \left( \frac{\kappa T}{n^2 \Lambda} \right)^{1/2}
\end{equation}

\noindent where $\kappa$ is the thermal conduction coefficient.

In order to include the effect of thermal conduction, the term 
$\nabla \left(\kappa \nabla T \right) / \rho C_v$ has to be added to the
right-hand side of equation (3). The linearized pressure equation (12) is then

\begin{equation}
\frac{\partial}{\partial \tau} \frac{ P_1}{P_0}= 
- \Gamma \sum_{j=1} ^3 V_j - \frac{1}{1-(1-\beta)\tau} 
\left((2-\beta)\frac{\rho_1}{\rho_0} - (1-\beta)\frac{P_1}{P_0} \right)
- \frac{\kappa}{\rho_0 C_v} \frac{T_1}{T_0} k^2 \tau_c(0) .
\end{equation}

In the isobaric regime with $P_1/P_0 \approx 0$ and $T_1/T_0 \approx - \rho_1/\rho_0$
thermal conduction will become important if

\begin{equation}
\frac{\kappa}{\rho_0 C_v} k^2 \tau_c(0) \geq \frac{2-\beta}{1-(1-\beta)\tau}
\end{equation}

\noindent In the early stages of cooling ($\tau << 1$) fluctuations will therefore be erased by thermal
conduction if their wavelenghts are

\begin{equation}
\lambda \leq \lambda_{\kappa} = \frac{2 \pi}{(2-\beta )^{1/2}} \lambda_F.
\end{equation}

Figure 7 shows the evolution of an initially isobaric, 1-dimensional fluctuation
with a ratio of cooling- to sound crossing time K=200 and $\beta = 0$. The 1-dimensional,
non-linear hydrodynamical equations (1) - (3) are solved numerically, including thermal 
conduction. The solid line shows the evolution of the density contrast $\rho_1/\rho_0$ 
as predicted by the analytical model (equation 22) which is in excellent agreement
with the numerical result (filled points) for $\lambda > \lambda_{\kappa}$. For
$\lambda = \lambda_{\kappa}$ (upper dashed line) conductive heating is non-negligible anymore
and the fluctuation grows less fast. For $\lambda < \lambda_{\kappa}$ the growth of fluctuations
is suppressed by thermal conduction.

In summary, thermal conduction can play a significant role in regulating the
break-up of a radiatively cooling gaseous medium. Small scale substructure can only
emerge in a limited wavelength regime which is given by 

\begin{equation}
\lambda_{\kappa} \leq \lambda \leq \lambda_{crit}
\end{equation}

\noindent The growth of perturbations is completely suppressed if
$\lambda_{crit} < \lambda_{\kappa}$ for all perturbations.

The dotted lines in figure 8a show $\lambda_{crit}$ for fluctuations with 
initial overdensities $\log (\rho_a/\rho_0) = -3,-2,-1$, the solid line shows 
$\lambda_{\kappa}$. A cooling function assuming
collisional equilibrium ionization (Spitzer 1978, Dalgarno \& McCray 1972) 
and a realistic conduction coefficient
(Ferrara \& Shchekinov 1993) has been adopted. The dashed line shows
the mean free path (Cowie \& McKee 1977)

\begin{equation}
\lambda_e \approx 10^4 \left( \frac{T}{K} \right)^2 \left( \frac{cm^{-3}}{n} \right) cm
\end{equation}

\noindent for electron energy exchange. Note that for a given temperature the ratios
$\lambda_{\kappa} / \lambda_{crit} / \lambda_e$ are independent of pressure.

The classical thermal conductivity is based on the assumption that $\lambda_e$ is short
compared to the temperature scale height $h_T \approx \lambda/(T_1/T_0) > \lambda_{crit}$.
Otherwise, the heat flux q is saturated (Cowie \& McKee 1977) 
and no longer equal to $q = -\kappa \nabla T$. Indeed, figure 8a shows that
linear fluctuations in astrophysical plasmas will in general lie in the non-saturated regime
($\lambda_{crit} > \lambda_e$) for  initial amplitudes $\rho_a/\rho_0 \geq 0.001$.

However, we also find that in general
$\lambda_{\kappa} > \lambda_{crit}$ for perturbations with amplitudes
$\log (\rho_a/\rho_0) \leq -2$. This implies that a cooling instability, resulting from
linear density perturbations will in general be suppressed by thermal conduction.

\subsection{Thermal conduction, including magnetic fields}

The interstellar medium is in general penetrated by magnetic fields. In most situations the
electron mean free path $\lambda_e$ is large compared to the length
scale at which the resistive destruction of the magnetic field is significant.
A tangled magnetic field can then develop, concentrated on scales $l_B$ which are
smaller than $\lambda_e$ (Chandran \& Cowley 1998). When the gyroradius

\begin{equation}
a = \frac{v_T m_e c}{e B} \approx 2.2 \times 10^8 \sqrt{\frac{T}{10^8K}} 
\left( \frac{\mu G}{B} \right) cm
\end{equation}

\noindent of thermal electrons with typical velocities $v_T = (kT/m_e)^{1/2}$ is much
smaller than $l_B$ or $\lambda_e$ the magnetic field controls the motion of
individual electrons. This condition is satisfied in many astrophysical plasmas
even if the magnetic field is too weak to be hydrodynamically important.

If $a$ is small compared to the length scale $\lambda$ of a fluctuation, heat is
conducted according to the classcal thermal conduction equation (Spitzer 1962), however
with a thermal conductivity $\kappa_B$ which is reduced from the classical
Spitzer value $\kappa$ as a result of the tangled magnetic
field by (Chandran \& Cowley 1998)

\begin{equation}
\kappa_B \approx \frac{0.1}{\ln (l_B/a)} \kappa .
\end{equation}

If we normalize B to its value for magnetic-to-thermal energy equipartition
$B_T = (24 \pi \rho R_g T)^{1/2}$ we find

\begin{equation}
\ln \left(\frac{l_B}{a} \right) = -3.1 + \ln \left( \frac{B}{B_T} \right)
+ 2 \ln \left( \frac{T}{K} \right) - 0.5 \ln \left( \frac{n}{cm^{-3}} \right) + 
\ln \left( \frac{l_B}{\lambda_e} \right) .
\end{equation}

For weak magnetic fields ($B \approx 0.01 B_T$)
and length scales $\l_B$ of order the electron mean free path
equation 40 leads to  $\ln (l_B/a) \geq 10$, by this reducing $\kappa_B$ by two
orders of magnitudes and the length scale of thermal conduction
by one order of magnitude.
As an example, the shaded area in figure 8b shows the wavelength regime 
$\lambda_{\kappa} \leq \lambda \leq \lambda_{crit}$ where linear fluctuations
with $\rho_a/\rho_0 = 10^{-2}$ could grow as a result of cooling,
assuming a pressure of $P/k_B = 10^3$ K cm$^{-3}$ and $l_B = 0.1 \lambda_e$.
The presence of a weak magnetic field can 
suppress thermal conduction efficiently, allowing small scale structure
with wavelengths $\lambda \approx 10^{-2}$ pc to emerge
as a result of cooling.

\section{The importance of heating and the emergence of a stable
two-phase medium.}

The calculations in sections 3 and 4 showed that cooling gas clouds with small
initial perturbations break up on a critical wave length
$\lambda_{crit}$ below which over density first becomes nonlinear.  If
the initial amplitude is a decreasing function of $\lambda$, the
clouds would break up on the smallest length where the local
radiative cooling law remains valid and fluctuations are not destroyed by conduction.
But, if the initial amplitude
is an increasing function of the wavelength,
\begin{equation}
\lambda_{crit} = \frac{2 \pi \tau{_c}(0)}{K_{crit}} 
\sqrt{\frac{R_g T_0(0)}{\mu}}.
\label{a30}
\end{equation}
\noindent small perturbations on scales $\lambda < \lambda_{crit}$ are
erased when the gas accumulates in the center of fluctuations with
$\lambda = \lambda_{crit}$.  Perturbations with $\lambda >
\lambda_{crit}$ do not become nonlinear but they break up into
substructures with $\lambda = \lambda_{crit}$.  

After a cooling time the dense, cold, non-linear perturbations are
embedded in a warmer, diffuse environment (see Fig. 9).  However, 
the example in figure 9 shows that the
cooling timescale of the inter-clump gas remains short compared to the
initial cooling timescale.  This gas therefore cools to a
ground state temperature $T_{min}$ shortly after $\tau =
1$. Subsequently the reversed pressure gradient would remove the
fluctuations unless they are gravitationally bound.

In order to maintain a stable two-phase medium a heating term must be
included (Field et al. 1969). Here we assume a power law
dependence of the heating rate

\begin{equation}
\Gamma_h = \Gamma_0 \rho^{\gamma}
\end{equation}

\noindent where $\Gamma_0$ and $\gamma$ are constants. In general, the
size of the whole cooling region is large compared to $\lambda_{crit}$
such that it cannot establish pressure equilibrium with the
surrounding confining medium during a cooling timescale. In this case,
the region cools isochorically and breaks up into
substructures on scales of $\lambda_{crit}$ before establishing
pressure equilibrium with the environment.  If the average gas density
$\rho_0$ is smaller than a critical value

\begin{equation} \rho_{\Gamma} = \left( \frac{\Gamma_0}{\Lambda_0}
T^{-\beta} \right)^{\frac{1}{2-\gamma}} \end{equation}

\noindent heating would dominate everywhere and the region would adjust to
a thermal equilibrium state where heating is balanced against cooling.
If $\rho_0 > \rho_{\Gamma}$ cooling dominates and
the density fluctuations would grow and become non-linear as discussed
in the previous sections. Eventually, after a cooling time, the region
would break up into cold high-density condensations which are separated
by warm gas with densities $\rho_{min} << \rho_0$. If $\rho_{min} >
\rho_{\Gamma}$ this interclump medium would cool as shown in figure 9 and
the density fluctuations would be erased.  Figure 10 shows a situation
with $\rho_{min} < \rho_{\Gamma}$. Heating dominates in the interclump
region where the gas temperature and gas pressure rise, until pressure
equilibrium is established. A stable 2-phase medium has formed with
cold clouds of minimum temperature embedded in a hot interclump medium
with a temperature that is determined by the balance of cooling and
heating.

\section{Discussions}

The discussions in this paper focussed on the emergence of small scale
perturbations.  We have assumed the pre-existence of small
initial perturbations which is a reasonable assumption for
dynamically evolving systems like the interstellar medium in galaxies
or in galactic clusters.  We have limited our analysis to the
optically thin regime such that radiation transfer is solely due to
optically thin local radiative processes.  This approximation is
appropriate for the collapse of supercritical clouds where the effect
of a magnetic field is dominated by thermal processes.  Such a
situation may be particularly relevant for the formation of stellar
clusters and first generation stars in galaxies. Provided that the density
of the progenitor clouds is relatively small, the local cooling
approximation is adequate.  We also neglected the interaction and merging of
clumps.  These processes become important for the subsequent evolution
and they will be considered in subsequent papers.

In the context of our approximations, we have shown that thermal
instability can lead to the breakup of large clouds into cold, dense
clumps with a characteristic length scale which is given by
$\lambda_{crit}$ in eq. ({\ref{a30}) or by the smallest unstable
wavelength that is not erased by thermal conduction,
depending on whether the amplitude of the initial
perturbation is an increasing or decreasing function of wavelength.
For linear perturbations with overdensities $\rho_a/\rho_0 \approx 0.01$
the critical wavelength lies in the regime of $10^{-3}$ pc to $10^{-1}$ pc,
depending on the initial temperature.
The emergence of small scale dense subcondensations is equivalent to
fragmentation. As in a thermally unstable region the cooling
timescale is shorter than the dynamical timescale,
gravity has no time to play an important role during this fragmentation
process.  $\lambda_{crit}$ may be either
smaller or bigger than the Jeans' length. In the latter case
gravity becomes important eventually. In general however,
thermally induced fragmentation of clouds with small initial
density fluctuations proceeds the onset of gravitational
instability of their individual clumps.  

In our analyses, we adopted an idealized power-law cooling function.  In
reality, the cooling efficiency would terminate when the main cooling
agents reach their ground state or establish an equilibrium with some
external heating source. The latter is necessary for the clouds to 
attain a two-phase medium.  Interaction between these two phases
may determine the pressure, density and infall rate of the cloud complex
as well as the dynamical evolution and size distribution of cloudlets and sub condensations.
The analysis of this interaction will be presented elsewhere.

\newpage

\acknowledgements

We would like to thank the referee, Andrea Ferrara, for helpful suggestions
and for pointing out the importance of thermal conduction.
A. Burkert thanks the staff of UCO/Lick Observatory for the hospitality
during his visits. We thank S.D. Murray for useful conversation.
This research was supported by NASA through NAG5-3056, and an astrophysics
theory program which supports a joint Center for Star Formation
Studies at NASA-Ames Research Center, UC Berkeley, and UC Santa Cruz.

\newpage

\begin{figure}
\centerline{\psfig{figure=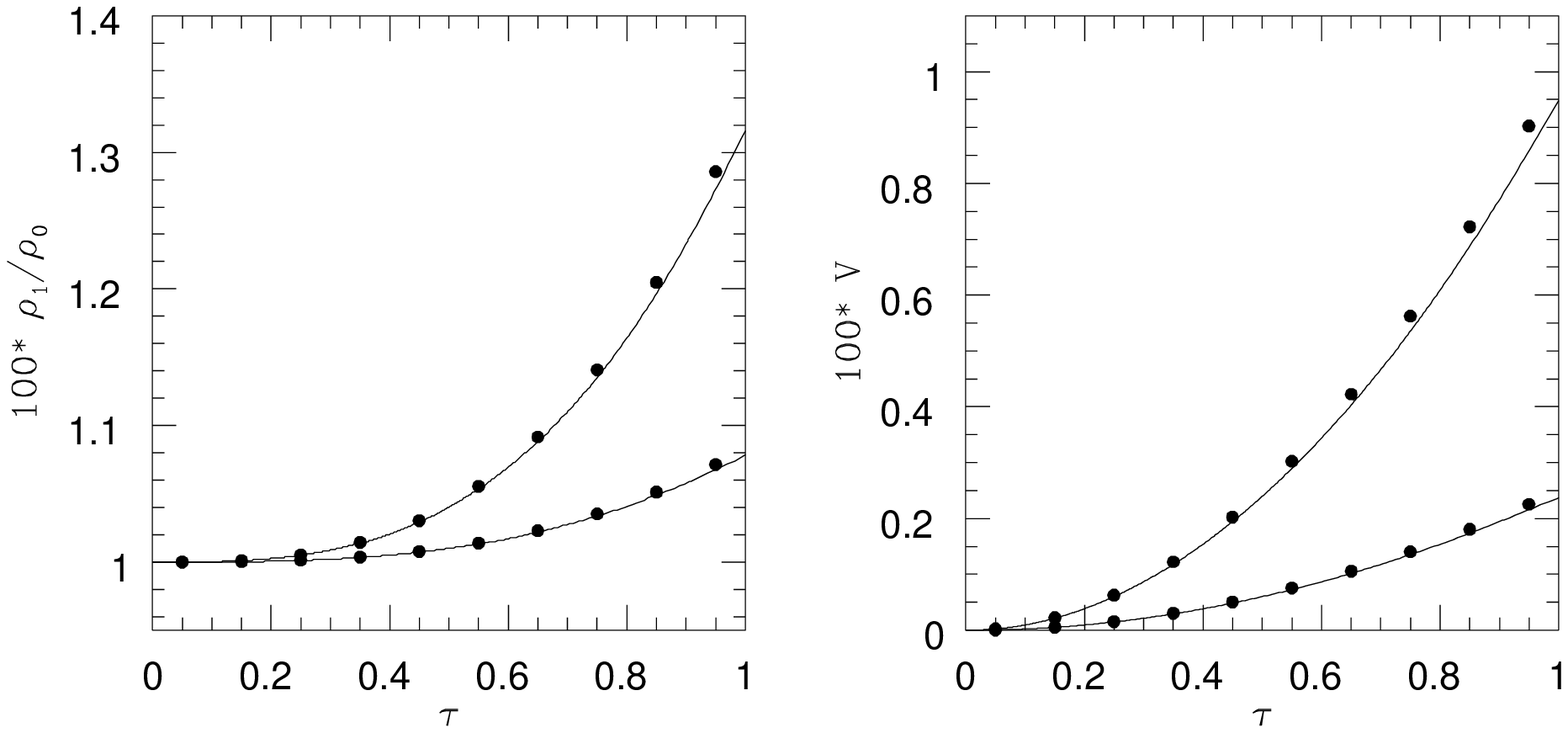,height=16cm}}
\caption {The solid lines show the numerical results of the
evolution of the overdensity
(left panel) and dimensionless velocity (right panel) for almost
isochoric perturbations with K=1 (upper curves) and K=0.5 (lower curves).
A cooling law with $\beta = 0$ is adopted. The dots represent the analytical
solution (equations 13 and 14).}
\end{figure}

\newpage

\begin{figure}
\centerline{\psfig{figure=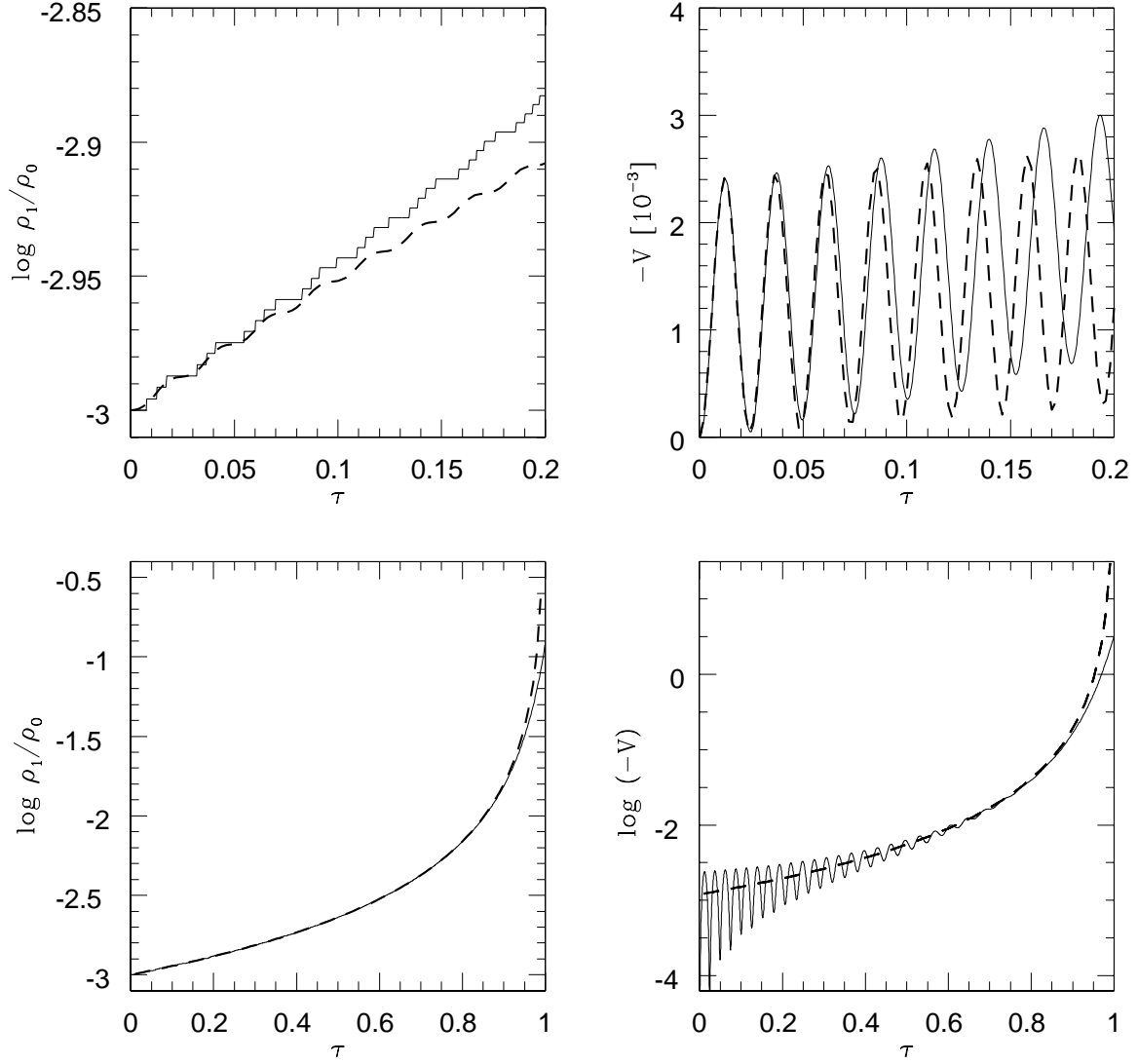,height=16cm}}
\caption {The growth of an initially isobaric density perturbation with 
K=200 is shown, adopting $\beta = 0$ and $\Gamma = 5/3$. The numerical results (solid lines) are compared
with the analytical solutions (dashed curves). The upper two panels show the early
evolution which is compared with the predictions of the linearized equations
(16) and (17) assuming $\omega = \sqrt{\Gamma}K$. The lower panels show the
evolution till $\tau=1$ which is compared with the results of the
nonlinear equations (22) and (23).}
\end{figure}

\newpage

\begin{figure}
\centerline{\psfig{figure=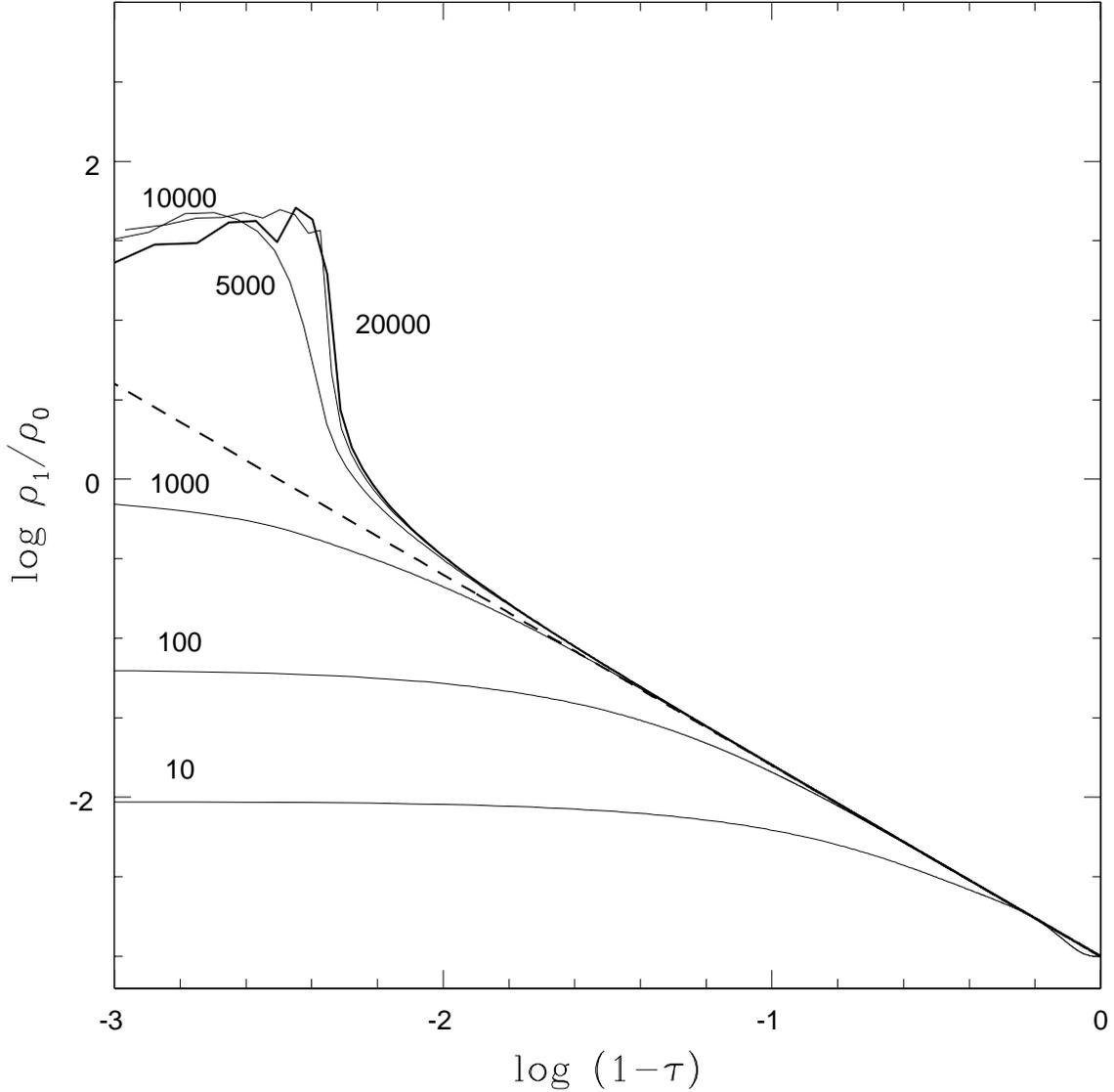,height=16cm}}
\caption{The growth of initially isobaric fluctuations is shown. The solid curves,
labeled by the adopted initial K-values of the fluctuations show the results of numerical
calculations for $\beta = 0$. The thick dashed line shows the evolution of 
a linear isobaric perturbation as predicted by equation (22). Fluctuations with small
values of K become isochoric and their density increase stalls. Above a critical value
of K fluctuations become nonlinear during their isobaric growth. These fluctuations
evolve faster than predicted by the linear approximation and  achieve very
large density contrasts on a timescale which is independent of K and shorter than
a cooling timescale.}
\end{figure}

\newpage

\begin{figure}
\centerline{\psfig{figure=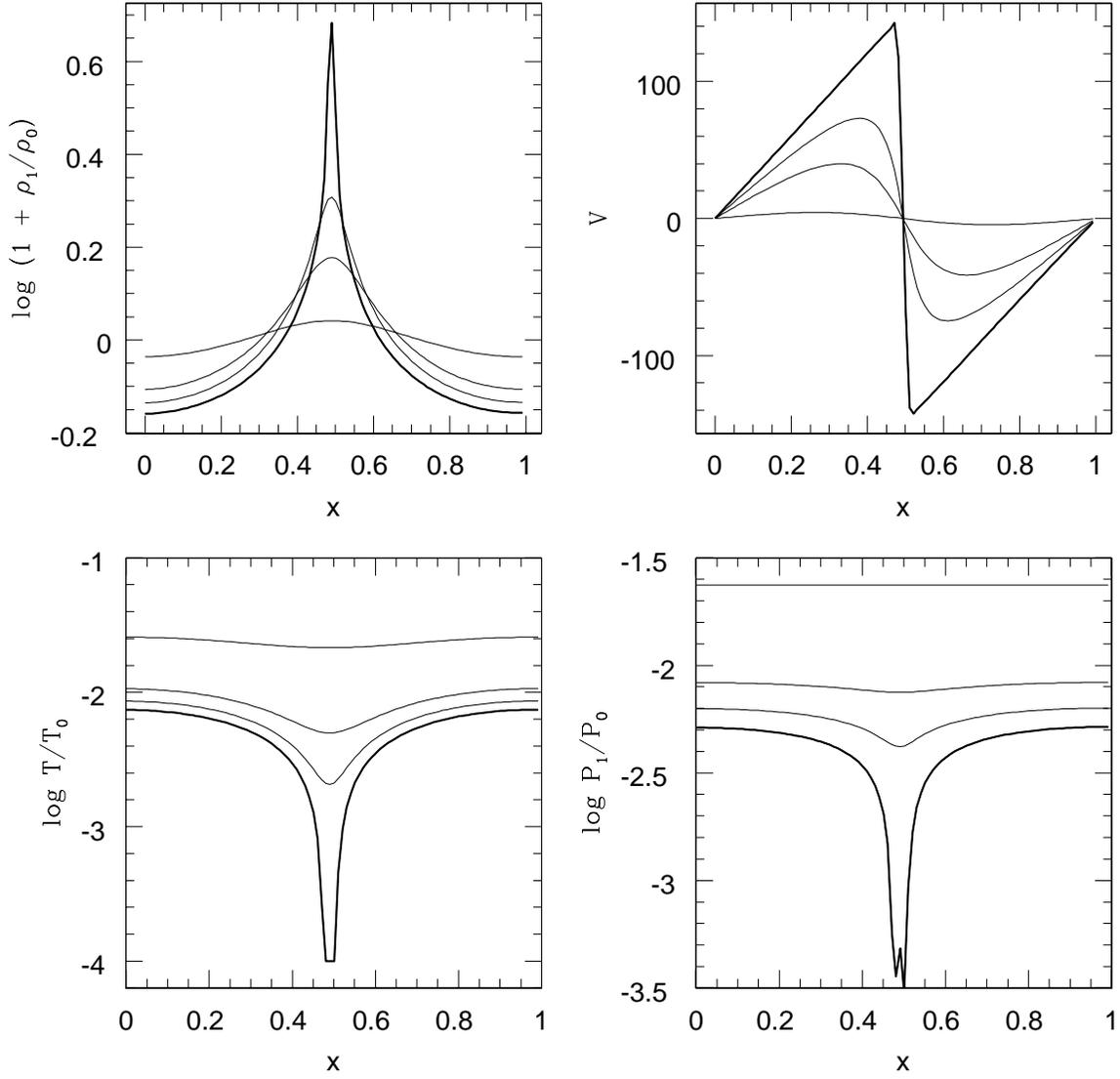,height=16cm}}
\caption{The evolution of overdensity $\rho_1/\rho_0$, dimensionless velocity V,
temperature $T_1/T_0$ and pressure $P_1/P_0$ of a nonlinear fluctuation with K=1000,
assuming $\beta=0$. Cooling is assumed to stop after the temperature has decreased
by 4 orders of magnitude. The evolutionary state of the system is shown
at the time when the overdensity is 0.1, 0.5, 1 and 5 times the initial density.}
\end{figure}

\newpage

\begin{figure}
\centerline{\psfig{figure=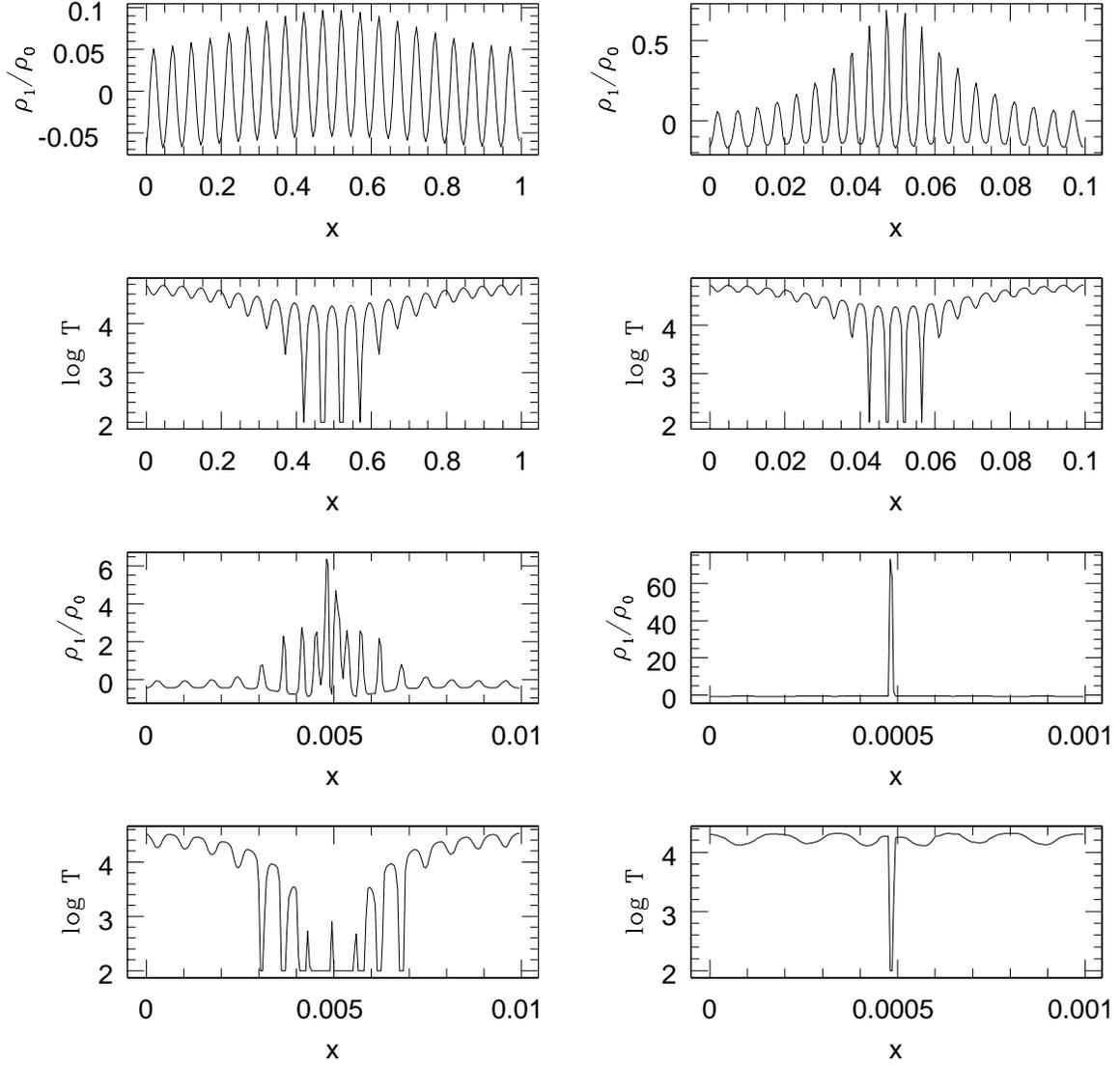,height=16cm}}
\caption {The figures show the overdensity and the temperature distribution
after a cooling time of regions with two interacting fluctuations with
wavelength ratios $\lambda_1/\lambda_0 = 20$ and amplitude 
ratios $\rho_{a,1}/\rho_{a,2} = 0.5$, adopting different values of K.
The x-coordinate is normalized to the wavelength of a perturbation with K=1.}
\end{figure}

\newpage

\begin{figure}
\centerline{\psfig{figure=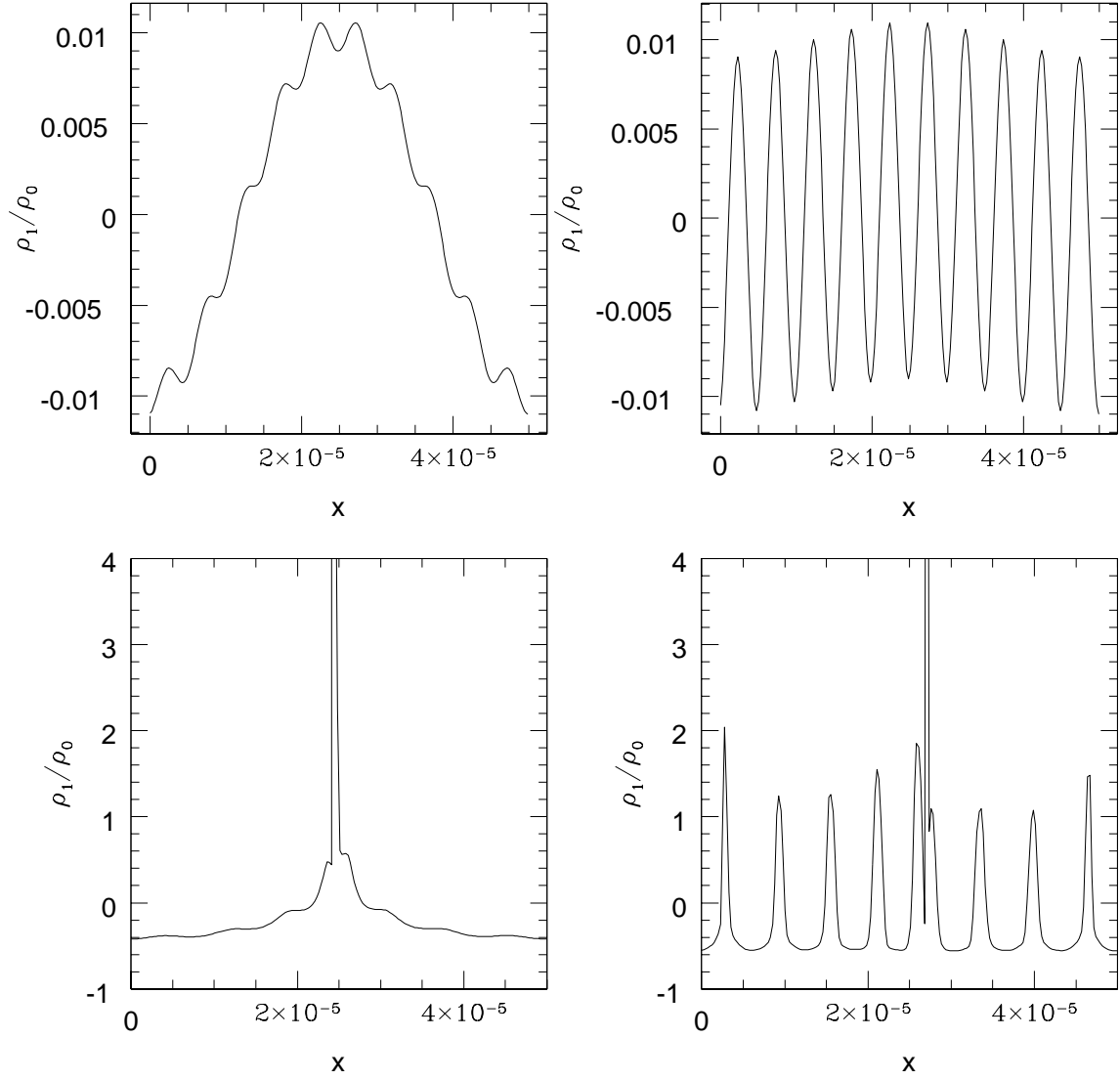,height=16cm}}
\caption{The left upper and lower panels show, respectively, 
the initial ($\tau = 0$) and 
final ($\tau = 1$) density distribution of two nonlinear interacting density perturbations 
with $\lambda_1/\lambda_0 = 10$ and $\eta=-1$. The initial and final density distributions
for interacting nonlinear fluctuations with $\eta=1$ are shown, respectively, in the
upper and lower right panels.}
\end{figure}

\newpage

\begin{figure}
\centerline{\psfig{figure=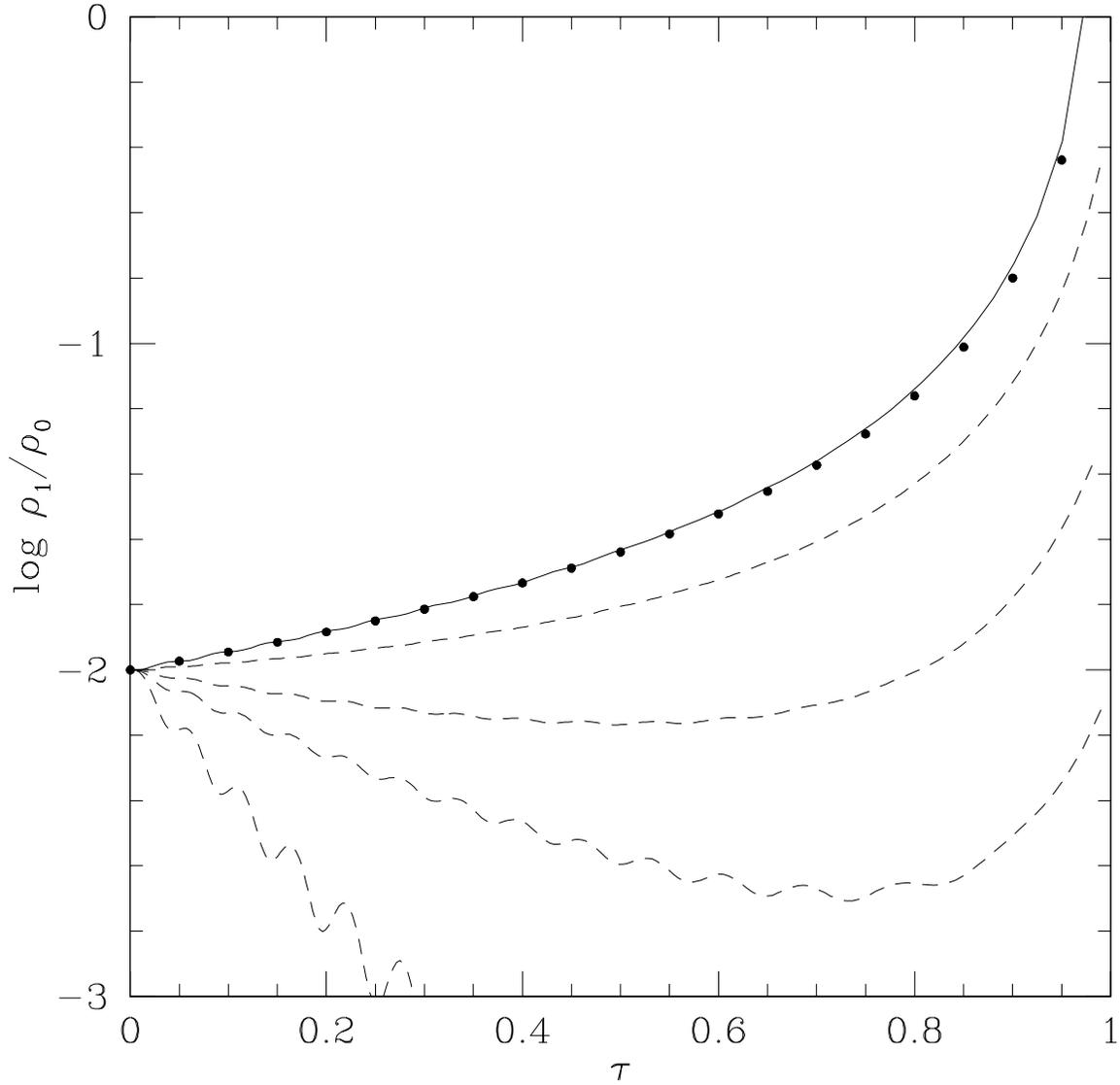,height=16cm}}
\caption{The effect of thermal conduction on the growth of a perturbation
with $\rho_a/\rho_0=0.01$ and $K=200$ is shown, adopting a cooling function with
$\beta = 0$. The solid thick line shows the theoretical prediction and the
filled points show the numerical result for $\lambda_{\kappa}=0$. The lower dashed
lines show the evolution of fluctuations with 
$\lambda=\lambda_{\kappa}$, $\lambda=0.75 \lambda_{\kappa}$, $\lambda=0.5 \lambda_{\kappa}$,
and $\lambda=0.25 \lambda_{\kappa}$, respectively.}
\end{figure}

\newpage

\begin{figure}
\centerline{\psfig{figure=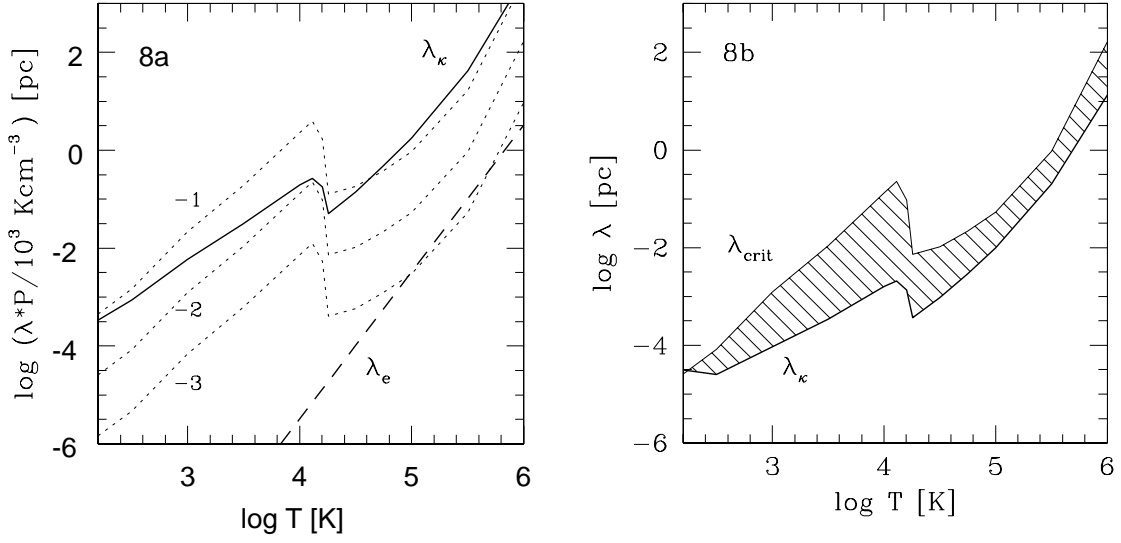,height=16cm}}
\caption{Fig. 8a shows several interesting lengthscales as function of
temperature, adopting a gas pressure of $P/k_B = n T = 10^3 K cm^{-3}$.
Solid line: $\lambda_{\kappa}$. Dashed lines: $\lambda_{crit}$ for density fluctuations
with $\log(\rho_a/\rho_0) = -1, -2, -3$, respectively. Dashed line: $\lambda_e$. 
The shaded area in Fig. 8b shows the wavelength regime $\lambda_{\kappa} \leq \lambda
\leq \lambda_{crit}$ where density perturbations with $\log (\rho_a/\rho_0) = -2$ would
be able to grow and become non-linear, adopting a gas pressure of $P/k_B = 10^3 K cm^{-3}$,
a magnetic field of $B = 0.01 B_T$ and a tangled magnetic field length scale of
$l_B = 0.1 \lambda_e$.}
\end{figure}

\newpage

\begin{figure}
\centerline{\psfig{figure=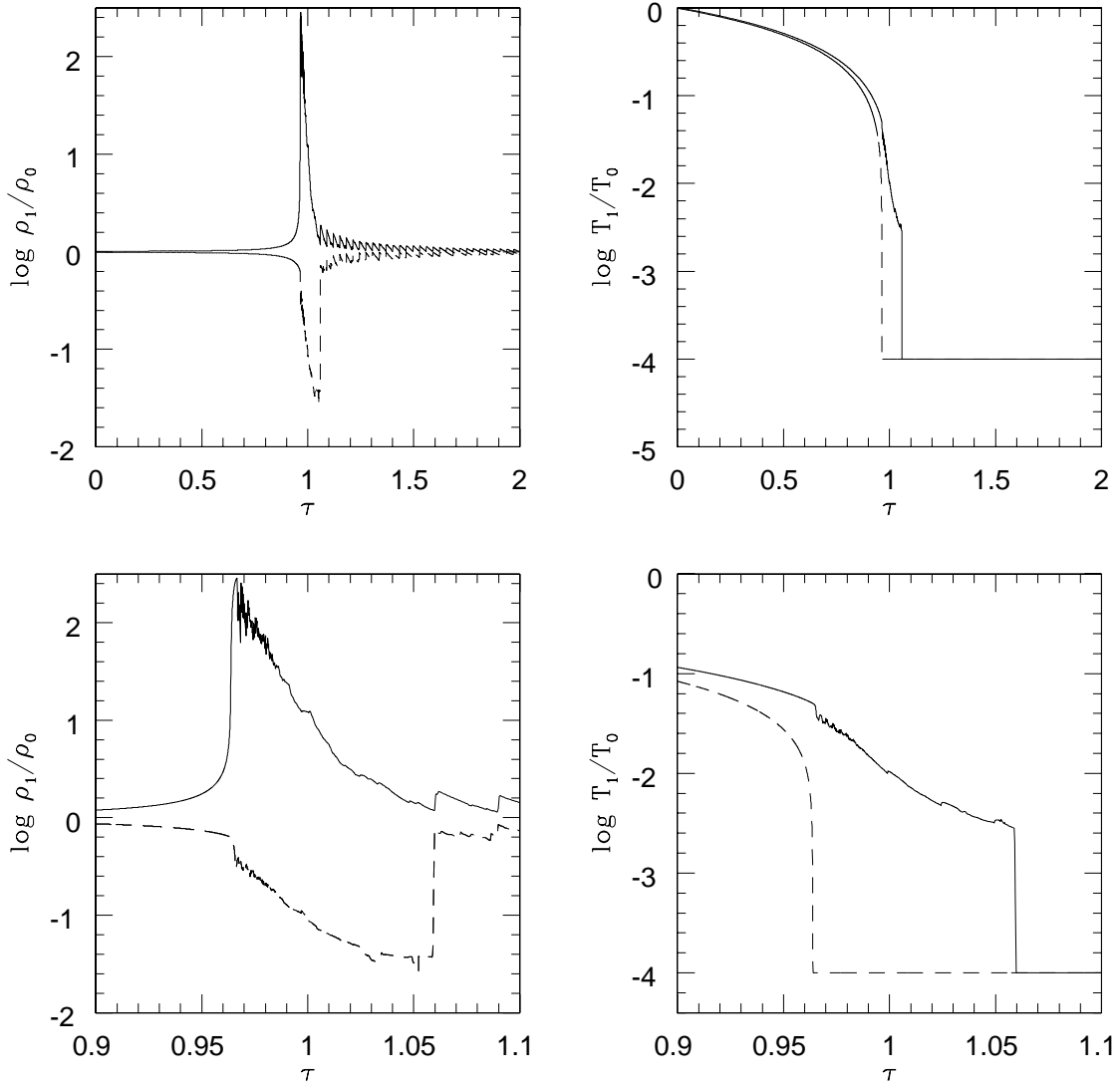,height=16cm}}
\caption{Evolution of a nonlinear perturbation with K=1000 and $\beta = 0$.
The upper panels show the evolution of the maximum density and temperature (solid line) 
and  of the minimum density and temperature (dashed lines)
for 2 cooling timescales. Shortly after one cooling timescale the low-density regions
cool down to the minimum temperature too and the reversed pressure gradient erases the
density contrast. The lower panels show the evolution of the density and temperature 
during the short epoch at $\tau \approx 1$ when a 2-phase medium has formed. }
\end{figure}

\newpage

\begin{figure}
\centerline{\psfig{figure=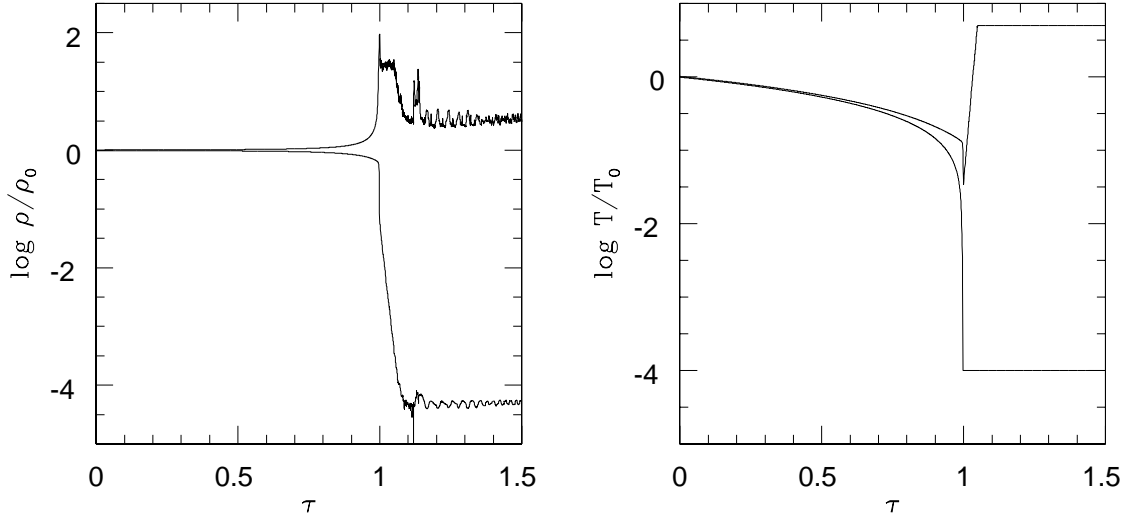,height=16cm}}
\caption{The evolution of the maximum and minimum density (left panel) and maximum
and minimum temperature (right panel) of a non-linear 
perturbation, including a cooling term with $\beta = 0$ and a heating term which
dominates for $\rho < \rho_{\Gamma}=0.1 \rho_0$, where $\rho_0$ is the initial average density. 
The density perturbation 
grows and becomes non-linear within a cooling timescale. As gas is pushed into 
the high-density regions,
the density in the inter-clump region decreases below the ciritical value where heating 
begins to dominate and the inter-clump gas heats up. 
A stable two-phase medium forms.}
\end{figure}

\end{document}